\documentclass[prd,preprint,superscriptaddress,showkeys,showpacs,
   preprintnumbers,amsmath,amssymb,aps,floatfix]{revtex4-1}
\usepackage{graphicx}
\usepackage{epsfig}
\usepackage{amsfonts}
\usepackage{amsmath}
\usepackage{amssymb}
\usepackage[sort&compress]{natbib}
\usepackage{dcolumn}
\usepackage{multirow}
\usepackage{bigstrut}
\usepackage{type1cm}
\usepackage{placeins}
\usepackage{rotating}
\usepackage{slashed}

\newcommand{\Ip}{\hat{\mathrm{I}}}

\begin{document}

\preprint{\today}

\title{Testing Lorentz invariance in orbital electron capture}

\author{K. K. Vos}
\author{H. W. Wilschut}
\author{R. G. E. Timmermans}

\affiliation{Van Swinderen Institute for Particle Physics and Gravity, 
University of Groningen, Nijenborgh 4, 9747 AG Groningen, The Netherlands}

\date{\today}
\vspace{3em}

\begin{abstract}
Searches for Lorentz violation were recently extended to the weak sector, in particular neutron and nuclear $\beta$ decay~\cite{Noo13a}.
From experiments on forbidden $\beta$-decay transitions strong limits in the range of $10^{-6}$-$10^{-8}$ were obtained on Lorentz-violating
components of the $W$-boson propagator~\cite{Noo13b}. In order to improve on these limits strong sources have to be considered. In this
Brief Report we study isotopes that undergo orbital electron capture and allow experiments at high decay rates and low dose. We derive the
expressions for the Lorentz-violating differential decay rate and discuss the options for competitive experiments and their required precision.
\end{abstract}

\maketitle

{\it Introduction.\/}---
Motivated by insights that Lorentz and CPT invariance can be violated in unifying theories of particle
physics and quantum gravity, a theoretical framework was developed in Refs.~\cite{Noo13a,Noo13b} to study Lorentz violation
in the weak gauge sector in neutron and (allowed and forbidden) nuclear $\beta$ decay. This approach, which parametrizes
Lorentz violation by adding a  complex tensor $\chi^{\mu\nu}$ to the Minkowski metric, includes a wide class of Lorentz-violating
effects, in particular contributions from a modified low-energy $W$-boson propagator $\langle W^{\mu+}W^{\nu -} \rangle =
-i (g^{\mu\nu}+\chi^{\mu\nu})/M_W^2$ or from a modified vertex $\Gamma^\mu=(g^{\mu\nu}+\chi^{\mu\nu})\gamma_\nu$.
Limits on Lorentz violation were subsequently extracted from experiments on allowed~\cite{Wil13,Mul13, Bod14} and
forbidden~\cite{Noo13b} $\beta$ decay, pion~\cite{Alt13b,Noo14}, kaon~\cite{Vos14}, and muon decay \cite{Noo15}.

The strongest bounds on components $\chi^{\mu\nu}$ were obtained~\cite{Noo13b} from forbidden-$\beta$-decay experiments~\cite{New76,Ull78}
and range from $10^{-6}$-$10^{-8}$ on different linear combinations. These bounds were translated in limits on parameters of the Standard
Model Extension~\cite{Col98}, which is the most general effective field theory for Lorentz and CPT violation at low energy. Specifically,
$\chi^{\mu\nu} = - k_{\phi\phi}^{\mu\nu} - i\,k_{\phi W}^{\mu\nu}/2g$ in terms of parameters in the Higgs and $W$-boson sector, where $g$
is the SU(2) electroweak coupling constant~\cite{Noo13a}. The resulting bounds on linear combinations of $k_{\phi\phi}^{\mu\nu}$ and
$k_{\phi W}^{\mu\nu}$ can be found in Ref.~\cite{Noo13b} and in the 2014 Data Tables in Ref.~\cite{Kos11}. The best bounds from allowed
$\beta$ decays are $\mathcal{O}(10^{-2})$ \cite{Mul13, Bod14} and from pion decay $\mathcal{O}(10^{-4})$ \cite{Alt13b}.

When seeking further improvement, one should realize that the bounds from forbidden $\beta$ decay benefited from the use of high-intensity
sources. Such strong $\beta$-decay sources, however, are hazardous because they have high disintegration rates (Bq) and high doses (Sv).
In this Brief Report we consider orbital electron capture~\cite{Bam77}, because the pertinent sources can give high decay rates at a low dose.
We first derive the theoretical expression for the differential decay rate including Lorentz violation. Next, we discuss the experimental possibilities
to constrain the various components $\chi^{\mu\nu}$. Finally, we explore which isotopes are suitable for a competitive measurement. 
We end with our conclusions.

{\it Decay rate.\/}---
 We consider allowed $K$-orbital electron capture~\cite{Kon66} mediated by $W$-boson exchange with a propagator that
 includes $\chi^{\mu\nu}$. We follow the notation and conventions of Ref.~\cite{Noo13a} ($\hbar=c=1$). The derivation of
the two-body capture decay rate is similar to the calculation of allowed $\beta$ decay~\cite{Noo13a}, but with the electron in
a bound state with binding energy $|E_K|$. Since Lorentz violation results in unique experimental signals, we restrict ourselves to the
allowed approximation with a nonrelativistic electron wave function with $\psi_e({\boldsymbol r}=0)=\sqrt{Z^3/(\pi a_0^3)}\,\chi_{s_e}$,
where $Z$ is the atomic number of the parent nucleus, $a_0=1/(\alpha m_e)$ is the Bohr radius, and $\chi_{s_e}$ is a Pauli spinor.
The neutrino is emitted with momentum ${\boldsymbol k}$ with $|{\boldsymbol k}|=E_\nu$ and the recoiling daughter nucleus has
momentum ${\boldsymbol p}_r=-{\boldsymbol k}$ and kinetic energy $T_r$. Because $E_\nu=Q-|E_K|-T_r\simeq Q$, the $Q$-value
of the reaction, the recoil energy is $T_r\simeq Q^2/(2M_r)$, which is typically 1-10 eV.

The differential decay rate is given by
\begin{equation}
dW = \frac{\delta(E_\nu - Q)}{(2\pi)^2 2 E_\nu} N_K \frac{1}{2} \sum_{s_e,s_\nu}|\mathcal{M}|^2 d^3k \ ,
\end{equation}
with $N_K=2$ the number of $K$-shell electrons. 
We define
$\xi = 2C_V^2\!\left\langle1\right\rangle^2+2\,C_A^2\!\left\langle\sigma\right\rangle^2$,
$x=2C_V^2\!\left\langle 1\right\rangle^2\!/\xi$,
$y=2C_VC_A\!\left\langle1\right\rangle\left\langle\sigma\right\rangle\!/\xi$, and
$z=1-x=2C_A^2\!\left\langle\sigma\right\rangle^2\!/\xi$,
where $C_V=G_F\!\cos\theta_C/\sqrt{2}$ and $C_A\simeq-1.27\,C_V$ are the vector and axial-vector coupling constants;
$M_F=\left\langle1\right\rangle$ and $M_{GT}=\left\langle\sigma\right\rangle$ are the Fermi and Gamow-Teller
reduced nuclear matrix elements. For a polarized source we find for the Lorentz-violating decay rate
\begin{eqnarray}
dW & = & dW^0 \left[ \left(1+B\,\hat{\boldsymbol k}\cdot\hat{\bf I} \right)\!/2 \right. \nonumber \\
&& \left. +\; t + {\boldsymbol w}_1\cdot\hat{\boldsymbol k} + \boldsymbol{w}_2\cdot {\bf \hat{I}}
               + T_1^{km}\,\Ip^k \Ip^m + T_2^{kj}\,\Ip^k \hat{k}^j + S_1^{kmj}\,\Ip^k \Ip^m \hat{k^j} \right] \ ,
\label{decayrate}
\end{eqnarray}
where $dW^0=(Z/a_0)^3E_\nu^2d\Omega_\nu\,\xi/(2\pi^3)$, $\hat{\boldsymbol k}={\boldsymbol k}/E_\nu$, and $\hat{\bf I}$
is the nuclear polarization axis. Latin indices run over the three spatial directions, with summation over repeated indices implied.
The Lorentz-violating tensors for electron capture read, in terms of the components $\chi^{\mu\nu}$,
\begin{subequations}
\begin{eqnarray}
t & = & (a-c/2)\,\chi_r^{00} \ , \\
w_1^j & = & -x\chi_r^{0j} - z(1+3\Lambda^{(2)}\!/2)(\tilde{\chi}_i^j-\chi_r^{j0})/3 \ , \label{w1} \\
w_2^k & = & -y\Lambda_z(\chi_r^{k0}-\chi_r^{0k})+z\Lambda^{(1)}\tilde{\chi}_i^k/2 \ , \label{w2} \\
T_1^{km} & = & 3c\,\chi_r^{km}\!/2 \ , \\
T_2^{kj} & = & A\chi_r^{00}\delta^{jk}\!/2 -z\Lambda^{(1)}(\chi_r^{jk}+\chi_i^{s0}\epsilon^{sjk})/2
                          +y\Lambda_z(\chi_r^{kj}+\chi_i^{0s}\epsilon^{sjk}) \ , \label{T2} \\
S_1^{kmj} & = & -3c\,(\chi_r^{k0}\delta^{mj}-\chi_i^{ms}\epsilon^{sjk})/2 \ .
\end{eqnarray}
\end{subequations}
The subscripts $r$ and $i$ denote the real and imaginary parts, respectively, of $\chi^{\mu\nu}=\chi^{\mu\nu}_r+i\chi^{\mu\nu}_i$, 
and $\tilde{\chi}^l = \epsilon^{lmk}\chi^{mk}$. The $V-A$ correlation coefficients~\cite{Jac57a,Her01,Sev06} that appear are
\begin{equation}
a = (4x-1)/3\ , \; c = z\Lambda^{(2)} \ , \;
A = z\Lambda^{(1)}-2y\Lambda_z \ , \; B = -z\Lambda^{(1)}-2y \Lambda_z \ . \\
\label{acAB}
\end{equation}
The angular-momentum coefficients $\Lambda^{(1)}$, $\Lambda^{(2)}$, and $\Lambda_z$ are given in the Appendix.
We absorbed a factor $\Lambda^{(2)}\!/3$ in $c$ and a factor $\left\langle m \right\rangle/j$ in $A$ and $B$~\cite{Noo13a}. 

Eq.~(\ref{decayrate}) reduces to the simple $V-A$ expression for the electron-capture decay rate when the Lorentz-violating
parameters are set to zero. In particular, the $B$ term in the first line of Eq.~(\ref{decayrate}) is the correlation between the spin
of the parent nucleus and the recoil direction of the daughter nucleus discussed in Refs.~\cite{Tre58,Fra58}. The second line
of Eq.~(\ref{decayrate}) gives Lorentz-violating, frame-dependent contributions to the decay rate.

{\it Observables.\/}---
From Eq.~\eqref{decayrate} we see that the possibilities to test Lorentz invariance in electron capture lie in measuring the decay rate as
function of either the nuclear polarization or the recoil momentum, or both. We restrict ourselves to dimension-four propagator corrections,
for which $\chi^{\mu\nu*}(p)=\chi^{\nu\mu}(-p)$ holds~\cite{Noo13a}. (The tensor $\chi^{\mu\nu}$ may contain higher-dimensional,
momentum-dependent terms, but such  terms are suppressed by at least one power of the $W$-boson mass.) Since $\chi$ is traceless,
this gives a total of 15 independent parameters, of which at present only $\chi_i^{0l}$ are unconstrained. The $\chi_r^{00}$ term will not
be considered, because it can only be accessed when comparing capture or $\beta$-decay rates between particles at rest and with a large
Lorentz boost factor $\gamma\gg 1$. In addition, we specialize to the suitable isotopes (identified below), which decay by Gamow-Teller
transitions, and for simplicity we assume that the source has vector polarization. This leaves
\begin{eqnarray}
dW & = & \frac{1}{2} dW^0 \left[ \left(1+B\,\hat{\boldsymbol k}\cdot\hat{\bf I} \right) \right. \nonumber \\
            && \left. -\; \left(\frac{2}{3}+\Lambda^{(2)}\right) (\tilde{\chi}_i^j-\chi_r^{j0}) \hat{k}^j + A \tilde{\chi}_i^k\, \Ip^k 
                - A \chi_r^{jk} \hat{k}^j \Ip^k - A \chi_i^{s0} (\hat{\boldsymbol k}\times\hat{ \bf I})^s \right] \ ,
\label{decayrate2}
\end{eqnarray}
where for pure Gamow-Teller decays $A=-B=\Lambda^{(1)}$. The different components $\chi^{\mu\nu}$
can be accessed by measuring asymmetries. We give three examples.

\begin{figure}[t]
\centering
\includegraphics[width=0.50\textwidth]{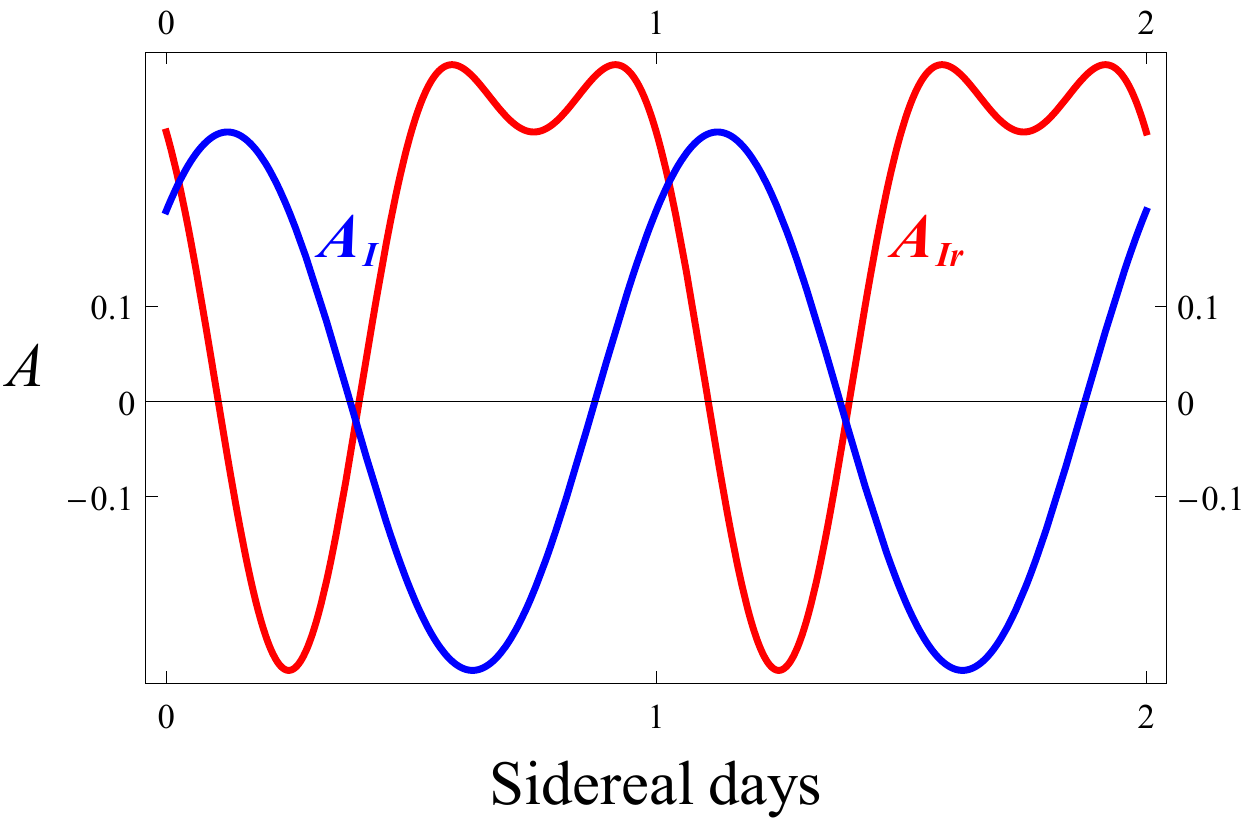}
\caption{The oscillation of the asymmetries in Eqs.~\eqref{eq:asym} and \eqref{eq:Ajr} as function of sidereal time, for $|\chi^{\mu\nu}|=0.1$,
$A=1$, and colatitude $\zeta=45^\circ$. To avoid a constant offset of the signal, we assumed polarization in the east-west ($\hat{y}$) direction.
For $\mathcal{A}_{Ir}$, $\hat{\boldsymbol k}$ was taken in the laboratory $\hat{z}$ direction, {\it i.e.} perpendicular to Earth's surface.}
\label{fig:asymmetries}
\end{figure}

($i$) $\chi_i^{jk}$ can be obtained from $\tilde{\chi}_i$, which can be measured from an asymmetry that depends
on the nuclear polarization, {\it viz.}
\begin{equation}\label{eq:asym}
\mathcal{A}_{I} = \frac{\tau^+ - \tau^-}{\tau^+ + \tau^-} = \frac{W^- - W^+}{W^+ + W^-} = -A \tilde{\chi}_i^k\, \Ip^k \ ,
\end{equation}
where $A$ contains the degree of polarization of the source and $\tau^{\pm}$ and $W^{\pm}$ are the lifetime and decay rate, respectively,
in two opposite polarization directions $\pm$. Such an experiment only requires to flip the spin of the sample and observe the change in
decay rate. For a discussion on using the direction of polarization to reduce systematic errors, see Ref.~\cite{Wil13}.
In general, the observables must be expressed in a standard inertial frame, for example the Sun-centered frame~\cite{Kos11}.
In the laboratory frame $\mathcal{A}_{I}$ will vary with $\Omega$, the angular rotation frequency of Earth, and the results depend on the
colatitude $\zeta$ of the site of the experiment~\cite{Noo13a}, {\it cf.} Fig.~\ref{fig:asymmetries}. In practice one searches for these variations
as function of sidereal time in order to isolate the Lorentz-violating signal and to reduce systematical errors. 

($ii$) The asymmetry of the recoil emission direction in an unpolarized sample is given by
\begin{equation}\label{eq:asymp}
\mathcal{A}_r =  \frac{ W(-{\hat{\boldsymbol k}})- W(\hat{\boldsymbol k}) }{W(-{\hat{\boldsymbol k}})+ W({\hat{\boldsymbol k}}) }
                            = \frac{2}{3}( \tilde{\chi}_i^j-\chi_r^{j0})  \hat{k}^j \ ,
\end{equation}
which requires measuring the recoil direction $\hat{\boldsymbol p}_r$ of the daughter nucleus. In this experiment it is necessary to rotate the
setup as a whole to isolate the Lorentz-violating signal and reduce systematic errors, as was done in Refs.~\cite{New76,Ull78}.
$\mathcal{A}_{I}$ and $\mathcal{A}_{r}$ have the same sidereal frequency, but they can differ in phase.

($iii$) When measurements of the recoil direction and polarization are combined, electron capture also offers the possibility to
constrain the parameters $\chi_i^{s0}$, for which no bounds have been set so far. Such an experiment should measure
$\chi_i^{s0}({\hat{\boldsymbol k}}\times\hat{\bf I})^s$, similar to a triple-correlation experiment to measure time-reversal violation
in $\beta$ decay. For example, the asymmetry
\begin{equation}\label{eq:Ajr}
\mathcal{A}_{Ir} = \frac{W(-{\hat{\boldsymbol k}})^+ W({\hat{\boldsymbol k}})^-
- W(-{\hat{\boldsymbol k}})^- W({\hat{\boldsymbol k}})^+}{W(-{\hat{\boldsymbol k} })^+ W({\hat{\boldsymbol k}})^-
+W(-{\hat{\boldsymbol k}})^- W({\hat{\boldsymbol k}})^+ } = 2A( \chi_r^{jk} + \chi_i^{s0} \epsilon^{sjk})  \hat{k}^j \Ip^k \ ,
\end{equation}
where $\hat{\boldsymbol k}$ is measured perpendicular to $\hat{\bf I}$, contains both $\chi_r^{jk}$ and $\chi_i^{s0}$.
The first term also produces sidereal oscillations with frequency $2\Omega$. The difference for the asymmetries in Eqs.~\eqref{eq:asym}
and \eqref{eq:Ajr} is illustrated in Fig.~\ref{fig:asymmetries}. 

{\it Isotopes.\/}---
The most stringent bounds found for a single component $\chi^{\mu\nu}$ so far are at a level of $\mathcal{O}(10^{-8})$, other
components are at least as small as $\mathcal{O}(10^{-6})$~\cite{Noo13b}. Most of the existing bounds concern linear combinations
of several components $\chi^{\mu\nu}$, so that cancellations are in principle possible. Assuming maximal fine-tuning, the best bound
for a real component is $\mathcal{O}(10^{-6})$ and for an imaginary term $\mathcal{O}(10^{-4})$~\cite{JacobPhD}. To achieve the highest
statistical relevance very strong sources should be considered. In order to reach $10^{-9}$ statistical accuracy a source with a strength
in the order of Curies (1 Curie-year $\simeq 10^{18}$ disintegrations) is required. For a high-statistics experiment a source
that decays exclusively by electron capture is attractive, because the  emission of ionizing radiation is  strongly reduced:
only X-ray emission and Auger electrons are involved. The most energetic radiation is due to internal bremsstrahlung, which
is suppressed by at least the fine-structure constant.

A list of possible isotopes is given in Table \ref{isotopes}. Which isotope is the most suitable depends on the  detection and production
method. The decay rate can be measured from the ionization current due to Auger processes and the shake-off of electrons that
follows capture. This requires that the radioactive isotopes are available as atoms, possibly in a buffer gas. In this way one can polarize
nuclei via optical pumping. The four isotopes for which this strategy is feasible are indicated in Table~\ref{isotopes}. To observe the
nuclear polarization, internal bremsstrahlung can be used, which is anisotropic with respect to the spin direction~\cite{Int71,Van88}.

\begin{table}[t]
\begin{tabular}{c|c|c|c|c} \hline
	\hline Isotope  &$t_{1/2}$ [s]&$Q$ [keV]  & $\,j^{\,\pi_i} \rightarrow j'^{\,\pi_f}$ &    \\ 
	\hline$^{37}$Ar  &$3.0\times 10^6$  &814 &$\frac{3}{2}^+\rightarrow\frac{3}{2}^+$&$\checkmark$   \\ 
	$^{49}$V  &$2.9\times 10^7$  &602 &$\frac{7}{2}^-\rightarrow\frac{7}{2}^-$ &   \\ 
	$^{55}$Fe  &$8.6\times 10^7$  &231 &$\frac{3}{2}^-\rightarrow\frac{5}{2}^-$ &   \\ 
	$^{71}$Ge  &$9.9\times 10^5$  &232 &$\frac{1}{2}^-\rightarrow\frac{3}{2}^-$  &  \\ 
	$^{131}$Cs  &$8.4\times 10^5$  &355 &$\frac{5}{2}^+\rightarrow\frac{3}{2}^+$ &$\checkmark$  \\ 
	$^{163}$Ho  &$1.4\times 10^{11}$  &2.6 &$\frac{7}{2}^-\rightarrow\frac{5}{2}^-$&$\checkmark$   \\ 
	$^{165}$Er  &$3.7\times 10^4$  &376 &$\frac{5}{2}^-\rightarrow\frac{3}{2}^-$ &$\checkmark$   \\ 
	$^{179}$Ta  &$5.7\times 10^7$  &106 &$\frac{7}{2}^+\rightarrow\frac{9}{2}^+$ &   \\ 
	\hline$^{53}$Mn  &$1.2\times 10^{14}$  &597 &$\frac{7}{2}^-\rightarrow\frac{3}{2}^-$&    \\ 
	$^{97}$Tc  &$1.3\times 10^{14}$  &320 &$\frac{9}{2}^+\rightarrow\frac{3}{2}^+$ &   \\ 
	$^{137}$La  &$1.9\times 10^{12}$  &621 &$\frac{7}{2}^+\rightarrow\frac{3}{2}^+$ &   \\ 
	$^{205}$Pb  &$5.5\times 10^{14}$  &51 &$\frac{5}{2}^-\rightarrow\frac{1}{2}^+$&   \\ 
	\hline \hline
	\end{tabular}
\caption{Isotopes that decay exclusively by orbital electron capture to a stable ground state. The top eight are relatively short-lived
               species that decay via allowed transitions, the bottom four are long-lived isotopes that undergo forbidden transitions.
               $^{163}$Ho is long-lived because of the very low $Q$-value. The isotopes check-marked in the last column can be polarized
               directly by optical pumping, or possibly also via an optically-pumped buffer gas. }\label{isotopes}
\end{table}

Because there are only four options we discuss the production of these isotopes separately:
\begin{itemize}
\item $^{37}$Ar can be produced in a reactor via the reaction $^{40}$Ca($n$,$\alpha$)$^{37}$Ar. A source of 35 mCi was produced
          from 0.4 g of CaCO$_3$ for a transient NMR experiment~\cite{PittPhD} to test the linearity of quantum mechanics~\cite{Wei89}. 
	An alternative method would be proton activation. A  cyclotron beam of 25 MeV  protons on $^{37}$Cl allows for a production
	of $10^7$ Bq/$\mu$Ah \cite{Kis75}, so that a source of one Curie can be produced well within a week.
\item The production of $^{131}$Cs was developed for brachytherapy. Neutron and proton activation are both options. Neutron activation
          is possible by using $^{130}$Ba \cite{Otu14}, and proton activation by using Xe or Ba isotopes, which gives a yield of
          $>10^7$ Bq/$\mu$Ah \cite{Cs131}. Commercial sources are available. Cs can be separated well from other radioactive by-products.
\item $^{163}$Ho is an isotope of interest for measuring the neutrino mass, and is studied by for instance the ECHo collaboration \cite{ECHo}.
	The production of Ho has been considered in detail \cite{Eng13}. The maximal production rate is projected to be about $10^4$ Bq/h, which
          is insufficient for a competitive measurement to test Lorentz invariance.
\item $^{165}$Er can be produced with a proton beam on a Ho target \cite{Er165} with a yield of $10^8$ Bq/$\mu$Ah. In view of its short half-life
         of 10 h, this is the only practical method. Although the production is sufficient, radioactive Ho is a by-product and Er cannot be separated
         effectively from Ho. 
\end{itemize}
We conclude that $^{37}$Ar and $^{131}$Cs are the only viable isotopes to obtain competitive values for $\chi^{\mu\nu}$. $^{37}$Ar has
the lowest ionizing yield and in this respect may be preferred. It can be polarized via a buffer gas, or by first exciting the atom into the
metastable state. In Ref.~\cite{PittPhD} the $^{37}$Ar nuclei were polarized by spin exchange with optically-pumped K atoms and a
nuclear polarization of 56\% was achieved.

The experimental apparatus for a measurement of $\mathcal{A}_{Ir}$ in Eq.~\eqref{eq:Ajr} could be based on that used to measure
the recoil in electron capture of
$^{37}$Ar, first used to verify the existence of neutrinos \cite{Rod52}. In particular, the crossed-field spectrometer developed at that
time \cite{Kof54} can be read with modern electronics and adapted to include polarization of $^{37}$Ar. It is necessary to detect
ionization currents instead of counting the recoils in order to accommodate the high event rate if one wants to aim for an accuracy
of $10^{-9}$. However, because there are no limits yet on $ \chi_i^{s0}$, such an experiment would immediately produce new results
with a much more modest effort, while allowing to investigate the systematic errors that will limit the ultimate high-statistics and
high-precision experiments.

{\it Conclusions.\/}---
We have explored the potential of orbital electron capture to put limits on Lorentz violation in $\beta$ decay. The limits set
in earlier work~\cite{Noo13b} are already so strong that high-intensity sources are required. A source with a strength of at least
one Curie that decays solely by electron capture may allow such experiments. Our survey limits the choice to $^{37}$Ar and
possibly $^{131}$Cs. The theoretical formalism for such experiments was developed, following Ref.~\cite{Noo13a}, in a form
applicable to any allowed electron-capture process. For one set of parameters quantifying Lorentz violation, no bound has been
obtained as yet. These can be accessed in an experiment that measures the recoil from the neutrino emitted from a polarized
nucleus, thus producing a new result while testing the viability of the suggested experimental program.

{\it Acknowledgments.\/}---
This investigation grew out of the {\em 33$^{\,rd}$ Solvay Workshop on Beta Decay Weak Interaction Studies in the Era of the 
LHC} (Brussels, September 3-5, 2014). We are grateful to  A. Young for discussions about electron-capture experiments and for
sending us Ref.~\cite{PittPhD}. We thank J. Noordmans for helpful discussions.  The research was supported by the Dutch
Stichting voor Fundamenteel Onderzoek der Materie (FOM) under Programmes 104 and 114.

{\it Appendix.\/}---
The angular-momentum coefficients in Eqs.~\eqref{w1}, \eqref{w2}, \eqref{T2}, and \eqref{acAB} are 
\begin{equation}
\Lambda^{(1)} =\left\{\begin{array}{ll}
\frac{\left\langle m \right\rangle}{j} & (j' = j-1) \\
\frac{\left\langle m \right\rangle}{j(j +1)} & (j' = j) \\
\frac{-\left\langle m \right\rangle}{j + 1} & (j'=j+1)
\end{array}
\right. \!\!\!\! \ , \;
\Lambda^{(2)} = \left\{\begin{array}{ll}
\frac{\left\langle m^2 \right\rangle - \frac{1}{3}j(j + 1)}{j(2j - 1)} & (j' = j-1) \\
\frac{-\left\langle m^2 \right\rangle + \frac{1}{3}j(j + 1)}{j(j + 1)} & (j' = j) \\
\frac{\left\langle m^2 \right\rangle - \frac{1}{3}j(j + 1)}{(j+1)(2j + 3)} & (j'=j+1)
\end{array}
\right. \!\!\!\! \ , \;
\Lambda_z = \frac{\left\langle m \right\rangle}{j}\sqrt{\frac{j}{j+1}}\delta_{jj'} \ ,
\end{equation}
where $j$ and $j'$ denote the initial and final nuclear spin, respectively, and $\left\langle m \right\rangle$ and
$\left\langle m^2 \right\rangle$ denote the incoherent average of $m$ and $m^2$ over the populations of the
states $m=-j,\ldots,j$. $\Lambda^{(2)}$ vanishes for unpolarized sources and for decays with $j=j'=\tfrac{1}{2}$.

\end{document}